\documentclass[aps,prc,reprint,superscriptaddress,numerical,showpacs,floatfix,nofootinbib]{revtex4-1}

\usepackage{graphicx}
\usepackage{subfigure}
\usepackage{dcolumn}
\usepackage{hyperref}

\begin{document}

\title{Highly charged ions in Penning traps, a new tool for resolving low lying isomeric states}

\author{A.T. Gallant}
\email[Corresponding author: ]{agallant@triumf.ca}
\affiliation{TRIUMF, 4004 Wesbrook Mall, Vancouver, British Columbia, V6T 2A3 Canada}
\affiliation{Department of Physics and Astronomy, University of British Columbia, Vancouver, British Columbia, V6T 1Z1 Canada}
\author{M. Brodeur}
\affiliation{TRIUMF, 4004 Wesbrook Mall, Vancouver, British Columbia, V6T 2A3 Canada}
\affiliation{Department of Physics and Astronomy, University of British Columbia, Vancouver, British Columbia, V6T 1Z1 Canada}
\affiliation{National Superconducting Cyclotron Laboratory, Michigan State University, East Lansing, Michigan 48824, USA}
\author{T. Brunner}
\affiliation{TRIUMF, 4004 Wesbrook Mall, Vancouver, British Columbia, V6T 2A3 Canada}
\affiliation{Physik Department E12, Technische Universit\"{a}t M\"{u}nchen, D-85748 Garching, Germany}
\author{U. Chowdhury}
\affiliation{TRIUMF, 4004 Wesbrook Mall, Vancouver, British Columbia, V6T 2A3 Canada}
\affiliation{Department of Physics and Astronomy, University of Manitoba, Winnipeg, Manitoba, R3T 2N2 Canada}
\author{S. Ettenauer}
\affiliation{TRIUMF, 4004 Wesbrook Mall, Vancouver, British Columbia, V6T 2A3 Canada}
\affiliation{Department of Physics and Astronomy, University of British Columbia, Vancouver, British Columbia, V6T 1Z1 Canada}
\author{V.V. Simon}
\affiliation{TRIUMF, 4004 Wesbrook Mall, Vancouver, British Columbia, V6T 2A3 Canada}
\affiliation{Max-Planck-Institut f\"{u}r Kernphysik, Saupfercheckweg 1, 69117, Heidelberg, Germany}
\affiliation{University of Heidelberg, 69117, Heidelberg, Germany}
\author{E. Man\'{e}}
\affiliation{TRIUMF, 4004 Wesbrook Mall, Vancouver, British Columbia, V6T 2A3 Canada}
\author{M.C. Simon}
\affiliation{TRIUMF, 4004 Wesbrook Mall, Vancouver, British Columbia, V6T 2A3 Canada}
\author{C. Andreoiu}
\affiliation{Department of Chemistry, Simon Fraser University, Burnaby, BC V5A 1S6, Canada}
\author{P. Delheij}
\affiliation{TRIUMF, 4004 Wesbrook Mall, Vancouver, British Columbia, V6T 2A3 Canada}
\author{G. Gwinner}
\affiliation{Department of Physics and Astronomy, University of Manitoba, Winnipeg, Manitoba, R3T 2N2 Canada}
\author{M. R. Pearson}
\affiliation{TRIUMF, 4004 Wesbrook Mall, Vancouver, British Columbia, V6T 2A3 Canada}
\author{R. Ringle}
\affiliation{National Superconducting Cyclotron Laboratory, Michigan State University, East Lansing, Michigan 48824, USA}
\author{J. Dilling}
\affiliation{TRIUMF, 4004 Wesbrook Mall, Vancouver, British Columbia, V6T 2A3 Canada}
\affiliation{Department of Physics and Astronomy, University of British Columbia, Vancouver, British Columbia, V6T 1Z1 Canada}

\date{\today} 

\begin{abstract}
The use of highly charged ions increases the precision and resolving power, in particular for short-lived species produced at on-line radio-isotope beam facilities, achievable with Penning trap mass spectrometers. This increase in resolving power provides a new and unique access to resolving low-lying long-lived ($T_{1/2} > 50$~ms) nuclear isomers. Recently, the $111.19(22)$~keV (determined from $\gamma$-ray spectroscopy) isomeric state in $^{78}$Rb has been resolved from the ground state, in a charge state of $q=8+$ with the TITAN Penning trap at the TRIUMF-ISAC facility. The excitation energy of the isomer was measured to be $108.7(6.4)$~keV above the ground state. The extracted masses for both the ground and isomeric states, and their difference, agree with the AME2003 and Nuclear Data Sheet values. This proof of principle measurement demonstrates the feasibility of using Penning trap mass spectrometers coupled to charge breeders to study nuclear isomers and opens a new route for isomer searches.
\end{abstract}

\pacs{21.10.Dr, 21.10.-k, 82.80.Qx, 82.80.Rt}

\maketitle

\section{Introduction}

Many advances in our understanding of nuclear structure have come from studying nuclei at the limits of existence, such as near the particle drip-lines. In very neutron-rich nuclei the discovery of nuclear halos~\cite{Tanihata1985}, neutron-skins~\cite{Tanihata1992,Suzuki1995}, and the emergence of new sub-shell closures and magic numbers~\cite{Sorlin2008602} have provided stringent tests for nuclear models. The emergence of new magic numbers gives rise to the possibility of low-lying long-lived ($T_{1/2} > 50$~ms) isomers, such as spin trap and K isomers~\cite{Walker1999}, in neutron-rich nuclei~\cite{PhysRevC.69.034317}. Nuclear isomers are of interest for a variety of reasons. In astrophysics isomers play a role in determining the abundances of the elements, since low lying isomers can be excited through thermal excitations in hot astrophysical environments~\cite{Aprahamian:2005fk}, \textit{e.g.} the abundance of $^{26}$Al~\cite{Illiadis2011}. Furthermore, isomers near new magic numbers in neutron-rich nuclei may play an important role in determining the r-process path. Isomers also provide sensitive tests for nuclear models. The life-time of a nuclear state depends on the overlap between the excited and ground states, a small difference in a model wavefunction can lead to widely different predictions for nuclear half-lives~\cite{Walker05}. Additionally, nuclear isomers are intimately tied to nuclear structure through nuclear shapes and high spin states~\cite{Walker1999,Walker05}.

The production rates of very neutron-rich nuclei tend to be very low, and the possibility of half-lives greater than $\approx 50$~ms limits the potential to study these isomers in traditional $\beta$ and $\gamma$-ray spectroscopy experiments. Another hindrance for traditional spectroscopy experiments is the potential for contaminants or large backgrounds~\cite{Blaum2004} from molecular beams of lighter nuclei. A viable alternative to study the excitation energies of these isomers, with lifetimes in the millisecond regime, is through mass measurements. Two mass measurement techniques that can be used to measure nuclear isomers are storage rings and Penning traps~\cite{Blaum2006}. The ability to study nuclear isomers has been demonstrated at the GSI storage ring, with the discovery of long-lived isomers in neutron rich hafnium and tantalum isotopes~\cite{Reed2010} and in the proton rich nucleus $^{125}$Ce~\cite{Sun2010,Sun2010_2}. Storage rings can measure isomer excitation energies as low as 100~keV~\cite{Sun2010_2}, however, the technique is limited to isomers with half-lives greater than several seconds due to the length of the cooling process.

Penning traps have been shown to be the most precise mass spectrometers for stable~\cite{Rainville2004} and unstable~\cite{Lunney2003} isotopes. This property along with the ability to perform measurements with low count rates, as low as a few ions per hour as demonstrated by SHIPTRAP~\cite{Block:2010fk}, allows nuclear structure to be studied in nuclides and isomers near the particle driplines. To facilitate the study of isomers in Penning traps, it is desirable that the time of flight resonances, obtained by the time-of-flight ion-cyclotron resonance technique~\cite{Konig1995}, be separated by more than one full-width half-maximum (FWHM) of the resonance line shape. The FWHM $\Delta \nu_{FWHM}$ is proportional to the inverse of the radio-frequency excitation time $T_{RF}$ as $ \Delta \nu_{FWHM} \propto T_{RF}^{-1}$~\cite{Konig1995}. The mass of the ion is obtained from the cyclotron frequency $\nu_{c}$ which is extracted from a fit of the theoretical line shape. The cyclotron frequency of an ion in a homogeneous magnetic field is related to its mass as,
\begin{equation}
	\nu_{c} = \frac{q B}{2 \pi m_{ion}},
	\label{eq:cyclotron_freq}
\end{equation}

\noindent
 where $q$ and $m_{ion}$ are the charge and mass of the ion of interest and $B$ is the magnetic field strength in the trap. Thus, the resolving power $\mathcal{R}$ required to separate an isomer from the ground state by one FHWM is,
\begin{equation}
	\mathcal{R} = \frac{m_{a}}{\Delta m_{a}} \propto \frac{q B T_{RF}}{2 \pi  m_{a}}
	\label{eq:precision_mass}
\end{equation}

\noindent
where $m_{a}$ is the atomic mass of the ground state of the nuclide of interest and $\Delta m_{a}$ is the difference in atomic masses of the isomer and ground state. In what follows resolved will be taken to mean separated by one FHWM of the line shape. Since the charge state and the excitation time in the Penning trap both enter into Eq.~(\ref{eq:precision_mass}) in the numerator a decrease in the excitation time can be compensated by the appropriate increase in the charge state. As a rule of thumb the excitation time is chosen according to the relation $T_{RF} \leq 3 \cdot T_{1/2}$.

The ability to resolve isomers in Penning traps was first demonstrated at ISOLTRAP, where the isomeric states of $^{84,78}$Rb were seen~\cite{Bollen1992}. If the hyperfine structure of a nucleus is known, then through state selective laser ionization it is possible to create isomerically pure beams from which the ground and isomer masses can be determined with a minimum of contamination~\cite{VanRoosbroeck2004,Weber2005}. If the hyperfine structure of the ground and isomeric states lie close to each other then it is possible to create pure beams by the use of an appropriate ion/Penning trap cleaning technique~\cite{Blaum2004,Eronen2008}. The shortest of these cleaning techniques still requires upwards of one-hundred milliseconds and cannot be used for nuclei and isomers with half-lives on the order of tens of milliseconds. Penning traps can also be used as a discovery machine with the first observation of new nuclides~\cite{Neidherr2009} or isomeric states~\cite{Block2008}, or to determine isomer excitation levels seen in decay spectroscopy~\cite{Schwarz2001,Weber2005,Ferrer2010}. 

\begin{figure}
	\begin{center}
		\includegraphics[width=8.6cm]{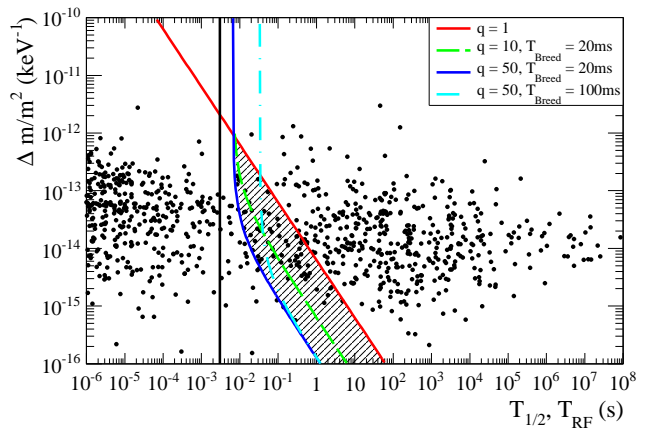}
	\end{center}
	\caption{\label{fig:resolve_allknownisomers}(color online) The normalized relative mass difference between an isomer and the ground state plotted against $T_{1/2}$ of the isomer. Superimposed is the available mass normalized resolving power, Eq.~\ref{eq:precision_mass}, for different $q$'s plotted against $T_{RF}$. Data are from Ref. \cite{NNDCWallet}. The vertical line at $3$~ms represents the shortest half-life that may be measured at TITAN. See text for details.}
\end{figure}

The increase in resolving power from charge breeding enables the resolution of low lying nuclear isomers in Penning traps. The dots in Fig.~\ref{fig:resolve_allknownisomers} show the normalized relative difference in mass $\Delta m / m^{2}$ for all known isomers with half-lives greater than 1~$\mu$s and excitation energies less than 9~MeV (868 in total) plotted against the half-life of the isomer. Superimposed as lines are calculations of the available mass normalized resolving power for a given $q$ and charge breeding time $T_{breed}$ assuming an excitation time of $3 \cdot T_{1/2} - T_{breed}$. The hatched area highlights the gain in resolving power, as these isomers could not be resolved as singly charged ions due to their short life-times. Isomers with half-lives much shorter than the breeding time required to reach a given charge state cannot be resolved. Any isomer to the right of the lines can, in principle, be resolved provided sufficient yield and a reasonable isomer to ground state production ratio. Ideally, in the absence of an isomer purification mechanism, both states would be delivered in equal amounts. A ratio considerably different than this would cause the resonance of the lesser produced species to be lost in the background produced by the primary species. In Fig.~\ref{fig:resolve_allknownisomers} isomers to the left of a line cannot be resolved with the chosen $q$. To demonstrate the improvement in the resolving power gained by using highly-charged ions to study nuclear isomers a mass measurement of $^{78}$Rb was performed in a charge state of $q = 8+$ and the reference ion was $^{85}$Rb$^{9+}$. For the first time the ground and isomeric states of $^{78}$Rb were resolved such that the minima were separated by more than one full-width half-maximum.

\section{The TITAN facility and experimental set-up}

\begin{figure}
	\begin{center}
		\includegraphics[width=8.6cm]{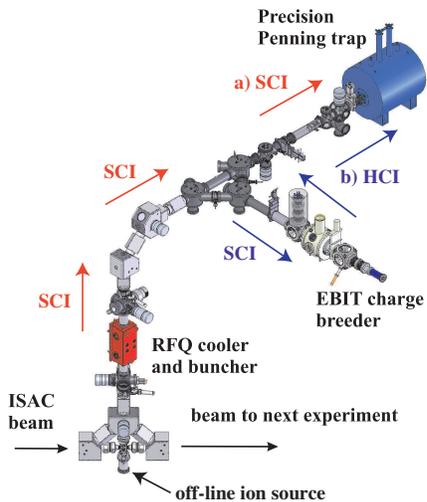}
	\end{center}
	\caption{\label{fig:TITANsetup}(color online) Illustration of the TITAN experimental set-up. For precision mass measurements singly charged ions delivered from ISAC or the offline source can (a) be injected directly into the MPET or (b) first sent to the EBIT for charge breeding and then sent to the MPET.}
\end{figure}

Triumf's Ion Trap for Atomic and Nuclear physics (TITAN)~\cite{Dilling2003,Dilling2006} is located in the ISAC hall of the TRIUMF laboratory. The TRIUMF-ISAC facility~\cite{Dombsky2000} produced the $^{78}$Rb beam by impinging a 500~MeV proton beam with  a current of up to 98~$\mu$A beam on a Nb production target. The produced radionuclides thermally diffuse out of the target where they are ionized and accelerated to an energy of 20 keV. The beam is cleaned of contaminants using a 120$^{\circ}$ dipole magnet with a resolving power of $m/\delta m \sim$3000 and delivered to TITAN. TITAN currently consists of three ion traps: a radio-frequency quadrupole (RFQ) cooler and buncher~\cite{Smith2007}, an electron beam ion trap (EBIT)~\cite{Lapierre2010} and a high precision measurement trap (MPET)~\cite{PhysRevC.80.044318, Brodeur2012}. A schematic of TITAN is shown in Fig. \ref{fig:TITANsetup}.

The RFQ is used to stop, thermalize, and bunch the beam delivered from ISAC in a helium buffer gas. This cooling decreases the energy spread of the beam which allows for a more efficient injection into either the MPET or EBIT. The beam is extracted from the RFQ with an energy of 2 keV. The cooling is of chief concern for the MPET since a large energy spread adversely affects the precision obtained in a mass measurement.

The MPET accepts beam either directly from the RFQ as singly charged ions (Fig. \ref{fig:TITANsetup}, path (a)) or from the EBIT as highly charged ions (Fig. \ref{fig:TITANsetup}, path (b)), and performs high precision mass measurements by utilizing the time-of-flight ion cyclotron resonance technique (see Ref.~\cite{Konig1995}). A magnetic field of $3.7$ T in the MPET is used to radially confine the ions while electrostatic fields are used to trap the ions axially. The general procedure for mass measurements with TITAN is outlined in \cite{PhysRevC.80.044318}. The TITAN system is a well established measurement facility for singly charged ions, in particular for very-short lived, neutron-rich light isotopes such as $^{6,8}$He~\cite{Brodeur2012b,Ryjkov2008}, $^{11}$Li~\cite{Smith2008}, and $^{11,12}$Be~\cite{Ringle2009,Ettenauer2010}. Recently, the first mass measurements of short-lived, highly charged ions were carried out at TITAN~\cite{Ettenauer2011}. 

\begin{figure}
	\begin{center}
		\includegraphics[width=8.6cm]{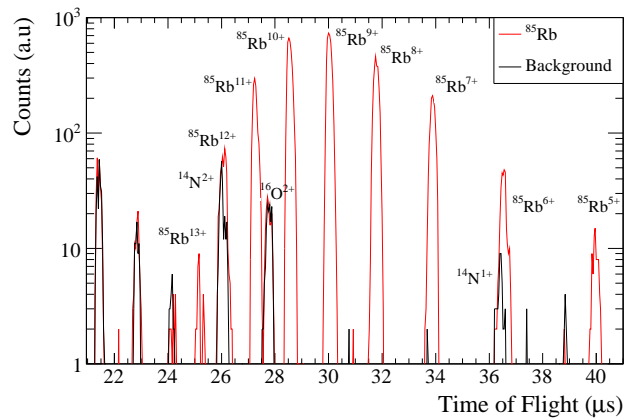}
	\end{center}
	\caption{\label{fig:Rb85MCP0}(color online) Time-of-flight spectra of charge bred ions from the EBIT detected on a micro-channel plate before the MPET. The red (black) curve shows the case of injection (no injection) of $^{85}$Rb into the EBIT.}
\end{figure}

The TITAN EBIT charge breeds the ions of interest stored initially as singly charged ions in the trap through successive impact ionization of the electron beam. The charge breeding process increases the ions' energy spread, which is estimated to be in the range of $10-100 \cdot q~\textnormal{eV}$~\cite{Ke2006}. In order to cool the charge bred beam a cooler Penning trap is currently being built and will use either electron or proton cooling to decrease the phase space of the highly charged ion bunch~\cite{Ryjkov2005,Simon10}. A full description of the TITAN EBIT can be found in Ref.~\cite{Lapierre2010,Gallant2010}. The ions are radially bound by a 3~T magnetic field and the space charge from the electron beam, and confined axially by a 100~V deep well. The injected ions and atoms of residual gas in the trapping region are bred to higher charge states and are extracted from the trap with an energy of $E_{\textnormal{kin}}\approx 1.9 \cdot q$ keV. The EBIT is typically opened for a short time, between a few hundred nanoseconds to a few microseconds, allowing for the selection of ions of different $m/q$'s by time-of-flight gating using a Bradbury Nielson gate~\cite{Brunner2011}. For this experiment the $^{78}$Rb beam was charge bred for 23~ms with an electron beam energy, relative to the trapped ions, of 2.5 keV and a current of $10$~mA, yielding a similar charge state distribution presented in Fig. \ref{fig:Rb85MCP0} for $^{85}$Rb. An $m/q$ ratio of $\approx 9.5$, which corresponds to ionic states of $^{85}$Rb$^{9+}$ and $^{78}$Rb$^{8+}$, was chosen due the cleanliness (lack of charge bred residual gas) of the TOF spectrum, hence, a clear separation of species could be achieved. 

\section{Data Analysis}

The quantity of interest in a Penning trap mass measurement is the cyclotron frequency $\nu_{c}$ of the trapped ions, and is related to the mass as shown in Eq. (\ref{eq:cyclotron_freq}). This is extracted from a fit of the theoretical line shape~\cite{Konig1995} to the time-of-flight resonance spectra. To eliminate major systematic shifts, such as those due to drifts of the magnetic field, the ratio $R$ of the cyclotron frequency of the ion of interest to that of a well known reference ion is taken in the combination such that the reference frequency is in the numerator,
\begin{equation}
	R = \frac{\nu_{c,ref}}{\nu_{c}}.
\end{equation}

\noindent
Taking the ratio in this way greatly simplifies the calculation of the covariance relations presented in Sec. (\ref{sec:systematics}). After averaging a number of frequency ratio measurements Eq.~(\ref{eq:cyclotron_freq}) can be used to determine the mass of the ion of interest relative to the mass of the reference ion,
\begin{equation}
	\label{eq:massfromratio}
	m_{a} = \frac{q}{q_{ref}} \bar{R} \left( m_{a,ref} - q_{ref} m_{e} + B_{e,ref} \right) + q m_{e} - B_{e},
\end{equation}

\noindent
where $\bar{R}$ is the weighted average of all frequency ratios and $B_{e}$ and $B_{e,ref}$ are the electron binding energies of the ion of interest and of the reference ion. The electron binding energies for Rb$^{8+,9+}$ are $\approx 500$~eV and $\approx 650$~eV \cite{Sansonetti2006}. The population of long-lived ionic metastable states that add significant amounts of energy, or equivalently, mass, to the system can be neglected, as the binding energies are relatively small.

\begin{figure}
    \begin{center}
	\includegraphics[width=8.6cm]{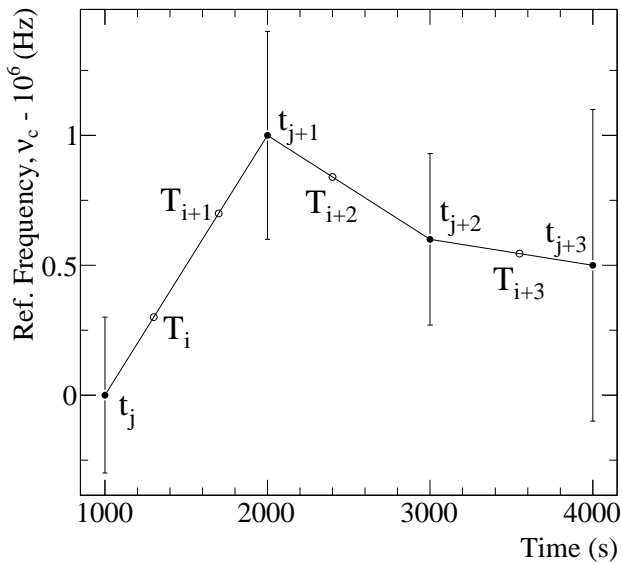}
    \end{center}
	\caption{\label{fig:covar_1}Illustration of the correlation introduced between adjacent frequency ratio measurements due to shared references. The filled circles ($t_{j}$'s) represent reference measurements of $\nu_{c,ref}$ and the open circles ($T_{i}$'s) show the interpolation of $\nu_{c,ref}$ to the center time of a measurement of an ion of interest. From the figure it is clear that $T_{i}$ is correlated with $T_{i+1}$, $T_{i}$ and $T_{i+1}$ with $T_{i+2}$, and $T_{i+2}$ with $T_{i+3}$ (details follow in the text).}
\end{figure}

\subsection{Time correlations between data sets}
\label{sec:systematics}

It is generally not practical to measure the reference frequency at the same time as the ion of interest, except for isomers or nuclides that lie very close in mass and are trapped quasi-simultaneously, as Coulomb interactions between the two stored species in the trap lead to systematic shifts in the measured frequencies~\cite{Bollen1992, Kellerbauer2003}. Therefore a linear interpolation between two frequency measurements, one taken immediately before and one immediately after the ion of interest must be made. As a time saving measure during an experiment two measurements of the ion of interest will often share a reference measurement introducing correlations between the frequency ratios. With the use of highly charged ions and the high level of precision that can be reached it is important to include these correlations when determining the final averaged frequency ratio. The relative statistical uncertainty of the cyclotron frequency in a measurement is related to the resolving power as $\delta m_{a} / m_{a} \propto \mathcal{R}^{-1} \cdot \sqrt{N}^{-1}$~\cite{Bollen20013} where $N$ is the number of detected ions. Here we present two cases shown in Fig. \ref{fig:covar_1}. First, the most likely case where two measurements share a reference measurement, and second, the case where several measurements occur between two reference measurements. In practice the second case does not occur since these data are generally summed. In the second case, the analysis with the time correlations and with the summed data will yield nearly the same result because the summed data implicitly includes time correlations between the frequency measurements. For the first case, the covariance relation between frequency ratios is,
\begin{equation}
\textnormal{covar}\left( R_{i}, R_{i+1} \right) = \frac{1}{\nu_{c,i} \nu_{c,i+1}} \left( \frac{T_{i} - t_{j}}{t_{j+1}-t_{j}} \right) \left( \frac{t_{j+2}-T_{i+1}}{t_{j+2}-t_{j+1}} \right) \sigma^{2}_{j+1}
\label{eq:covar_sameref}
\end{equation}

\noindent
where the $i$ and $i+1$ refer to the $i$th and $i$th$+1$ measurements of the ion of interest and the $j$'s refer to the reference measurements. For the second case, the covariance between frequency ratios that share both references is,
\begin{eqnarray}
\textnormal{covar}\left( R_{i}, R_{i+1} \right) & = & \frac{1}{\nu_{c,i} \nu_{c,i+1}} \left[ \left( \frac{t_{j+1}-T_{i}}{t_{j+1}-t_{j}} \right) \left( \frac{t_{j+1}-T_{i+1}}{t_{j+1}-t_{j}} \right) \sigma^{2}_{j} \nonumber \right. \\
&& \left. +\: \left( \frac{T_{i}-t_{j}}{t_{j+1}-t_{j}} \right) \left( \frac{T_{i+1}-t_{j}}{t_{j+1}-t_{j}} \right) \sigma^{2}_{j+1} \right].
\label{eq:covar_betweenref}
\end{eqnarray}

\noindent
Figure \ref{fig:covar_1} illustrates the relationship between the variables given in the above equations. In both cases the covariance is proportional to the variance of the reference measurements. It is desirable to measure the reference ion much more precisely than the ion of interest to reduce correlation effects, however, a trade-off must be made in order to maximize the statistics collected, and hence, the precision, for the ion of interest.

\subsection{Systematic errors/Uncertainties}

Several systematics must be taken into account. Systematics relating to misalignment between magnetic and trap axes,  electric field miscompensation, relativistic effects, etc., are minimized by choosing a reference ion which is close in $m/q$ to the ion of interest as these effects scale with the difference in the charge to mass ratio $\Delta(m/q)$ \cite{Brodeur2012}. To determine any potential shifts due to different $m/q$ effects between the ion of interest and the reference ion, a series of mass measurements on $^{85}$Rb$^{10,8+}$ and $^{87}$Rb$^{9+}$ using $^{85}$Rb$^{9+}$ as the reference were completed. The extracted masses all agree within 1$\sigma$ of the literature value. Although no shifts were observed to be conservative we take, as an upper limit on any systematic effects, a systematic uncertainty of 42 parts-per-billion (ppb) in the frequency ratio.

\begin{table*}[!]
	\caption{\label{tab:Rbfreqratios}Frequency ratios and mass excesses of $^{78,78n}$Rb$^{8+}$ measured relative to $^{85}$Rb$^{9+}$ with the total uncertainties. The error budget is summarized in Table \ref{tab:RbErrorBudget}.}
\begin{tabular*}{1.0\textwidth}{@{\extracolsep{\fill}} c c c c d}
	\hline
	\hline
	Isotope & $\nu_{c,ref}/\nu_{c}$ & $ME$ (keV) & $ME_{AME03}$ (keV) & \multicolumn{1}{c}{$ME - ME_{AME03}$ (keV)}\\
	\hline
	$^{78}$Rb$^{8+}$   & 1.032475265(99) & -66933.2(7.0) & -66936.228(7.452) & 3.0 \\
	$^{78n}$Rb$^{8+}$ & 1.032476806(60) & -66824.8(4.2) & -66825.038(7.455) & -0.2 \\
	\hline
	\hline
\end{tabular*}
\end{table*}

\begin{figure}
	\begin{center}
		\includegraphics[width=8.6cm]{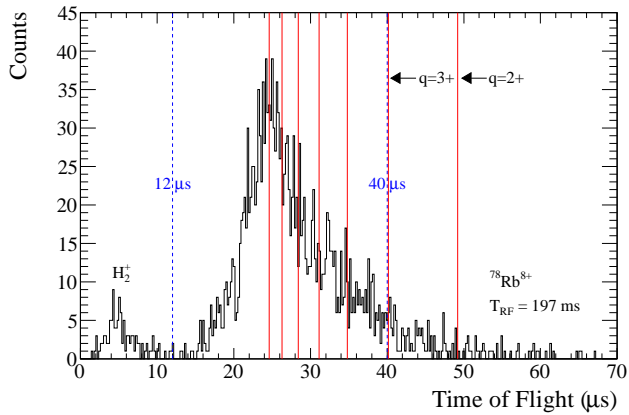}
	\end{center}
	\caption{\label{fig:TOFDistroWithChargeExchange}(color online) Time of flight spectrum of ions extracted from the MPET. See text for details.}
\end{figure}

A second systematic effect stems from the ambiguity in selecting the upper and lower time cuts on the time of flight spectrum. The ambiguity arises due to charge exchange processes in the trap. If an ion undergoes charge exchange with residual gas in the vacuum, these ions will manifest themselves as a long tail in the time of flight spectrum. In the present analysis the lower and upper levels were set at 12 and 40~$\mu$s, respectively. The lower level was set to 12~$\mu$s in order to maximize the number of on resonance ions while minimizing background counts from the nearby H$_{2}^{+}$ peak resulting from charge exchange in the trapping region.  Figure \ref{fig:TOFDistroWithChargeExchange} shows a typical time of flight spectrum for $^{78}$Rb$^{8+}$ which was trapped for 197~ms. The dashed-blue lines show the lower and upper time cuts while the solid-red lines show when, on average, $^{78}$Rb ions with different charge states would arrive. In order to determine the systematic effect $\bar{R}$ was determined for upper level time cuts of 30, 35, 40, 45 and 55~$\mu$s for both the ground and isomeric states. If the average frequency ratio determined at 40~$\mu$s for either case was an extremum the systematic effect was assigned to be the full range of the extracted $\bar{R}$'s, otherwise half of the range was assigned.

The last systematic effect results from ions with different $m/q$'s which are in the trap at the same time. For ions that are separated by more than the full-width half-maximum of the resonance curve the measured cyclotron frequency of both ions decreases from the nominal value~\cite{Bollen1992}. In order to eliminate this potential shift a count class analysis~\cite{Kellerbauer2003} was performed. For this the frequencies are determined as a function of the number of detected ions, hence the number of ions stored in the trap (count class). The cyclotron frequencies of the ground and isomeric states were extracted for different count classes and the frequencies were extrapolated to the detector efficiency of $0.6 \pm 0.2\%$. When one ion was detected after extraction on average about 1.7 ions were actually in the trap, thus, an extrapolation past unity was required. The count class analysis eliminated any correlations between the extracted fit parameters, so their correlations were not included in the analysis.

\subsection{Extracting isomeric excitation energies}

The excitation energy of the isomer can be extracted by using Eq.~(\ref{eq:massfromratio}), and, using the isomer as the reference, the difference in mass is

\begin{equation}
	m_{a,iso} - m_{a,gnd} = (1-\bar{R})(m_{a,iso} - q m_{e} + B_{e}).
\end{equation}

\noindent
Since the ground state and isomer represent a mass doublet, many of the systematic effects presented in the previous section cancel. However, to be conservative we include the maximum 42 ppb systematic shift presented in the previous section.

If other reference measurements were performed, it is also possible to extract the mass excess of each state as well as the energy difference of the isomeric state through a joint fit of all the frequency ratio pairs of the ground and isomeric states, making use of the full covariance matrix between all the frequency ratios using Eqs. \ref{eq:covar_sameref} and \ref{eq:covar_betweenref}. This process provides a better error estimate on the difference than achieved using simple error propagation methods. The weighted average of each frequency ratio, to be used in Eq. \ref{eq:massfromratio}, can be extracted from the same formalism and will yield a higher uncertainty than a weighted average assuming independent measurements since the correlations are positive. The technique used is fully described in Ref.~\cite{Valassi2003}.

\section{Results}

\begin{figure}
	\begin{center}
		\subfigure{ 
                  	\includegraphics[width=8.6cm]{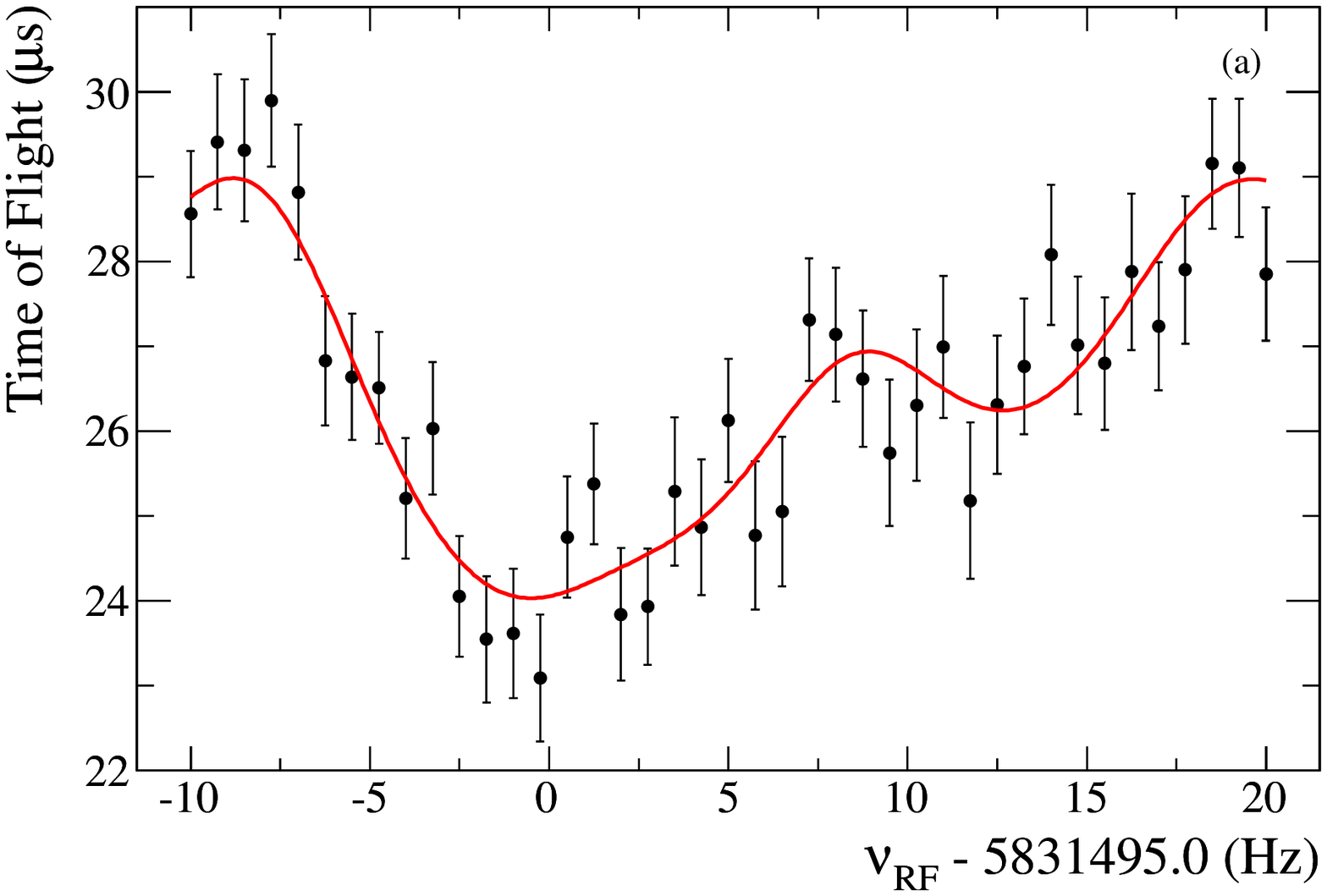}
		}
		\subfigure{
			\includegraphics[width=8.6cm]{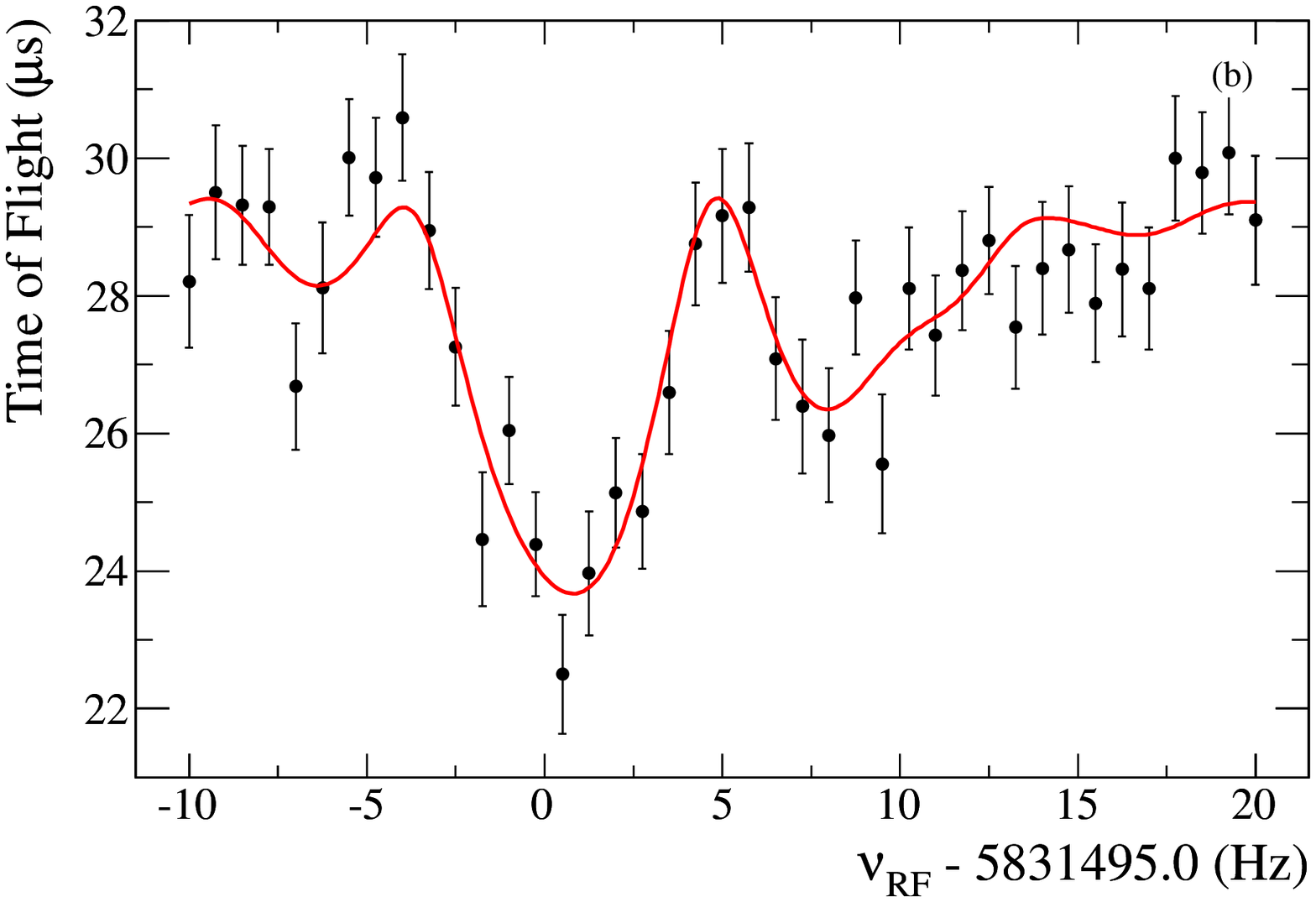}
		}
	\end{center}
	\caption{\label{fig:Rb78resonance}(color online) Resonance of $^{78}$Rb$^{8+}$ for quadrupole excitation times of (a) 97~ms and (b) 197~ms. The isomer resonance is on the left and the ground state resonance is on the right. The solid line is a fit of the theoretical line shape \cite{Konig1995} to the data. }
\end{figure}

\begin{table}
	\caption{\label{tab:RbErrorBudget}Error budget of the frequency ratios in ppb. ``Ground" and ``Isomer" indicate the measurements were done relative to $^{85}$Rb${^{9+}}$, while ``Doublet" indicates that the isomer was used as the reference ion.}
\begin{tabular}{c | c  d  d}
	\hline
	\hline
	Uncertainty (ppb) & Ground & \multicolumn{1}{c}{Isomer} & \multicolumn{1}{c}{Doublet} \\
	\hline
	Statistical			& 66.8	& 29.5	& 66.7	\\
	Time Correlations	& 24.1	& 24.5	& 0.0		\\
	TOF Spectrum Cut	& 55.0	& 20.0	& 38.0	\\
	Trap Systematics		& 42.0	& 42.0	& 42.0	\\
	\hline
	Total				& 99.0	& 60.0	& 87.0	\\
	\hline
	\hline
\end{tabular}
\end{table}

\begin{table}
	\caption{\label{tab:energylevel}Extracted excitation energy of the isomeric state. The error in parentheses is the statistical and time correlation error, while the error in braces is the total error. The time correlations in this measurement are of the second type (Eq. \ref{eq:covar_betweenref}).}
\begin{tabular}{c c}
	\hline
	\hline
	Method	& Excitation Energy (keV) \\
	\hline
	Time Correlation	& 108.4(4.8)\{7.6\} \\
	No Time Correlations & 108.5(5.1)\{7.8\} \\
	$^{78n}$Rb$^{8+}$ as reference	& 108.7(4.8)\{6.4\} \\
	\hline
	\hline
\end{tabular}
\end{table}

The frequency ratios for $^{78}$Rb$^{8+}$ (ground state) and $^{78n}$Rb$^{8+}$ (isomer\footnote{We use the notation introduced in the \MakeUppercase{Nubase}~\cite{Audi2003_2} evaluation where isomeric states are labeled by m, n, p, q, etc. in order of increasing excitation energy.}) ($T_{1/2} = 17.66(3) ~\textnormal{min},~5.74(3) ~\textnormal{min}$~\cite{NNDC}) measured relative to $^{85}$Rb$^{9+}$ are presented in Table \ref{tab:Rbfreqratios}. The error budget, showing the contributions from all systematics discussed previously, is presented in Table \ref{tab:RbErrorBudget}. Typical resonances for 97~ms and 197~ms excitation times are shown in Fig. \ref{fig:Rb78resonance}. Subtracting the literature excitation energy of the isomer from the measured mass excess and taking the weighted average of both measurements (at $T_{RF}=97,197$~ms), it was found that the mass excess for $^{78}$Rb is $-66935.3(3.6)$~keV. This is in excellent agreement with the AME03 value~\cite{Audi2003}.

Extracting the mass of the ground state was complicated by two factors. First, the ratio of isomeric to ground state nuclei delivered from ISAC was $N_{isomer}/N_{g.s.} \approx 2$. The difference in the number of ions weakens the time-of-flight resonance of the ground state leading to a larger uncertainty in the fitted frequency. Second, while the resonances are fully resolved in Fig. \ref{fig:Rb78resonance}(b), the ground state sits very near the first sideband of the isomer resonance leading to the ambiguity of what the fitting routine is extracting: The ground state center frequency or the position of the first sideband of the isomer? In our case the sideband of the isomer is half of the depth of the ground state resonance causing the isomer sideband to have an affect on the fitted ground state frequency.

For the TITAN set-up full resolution of the isomer and ground states is achieved with a quadrupole excitation time of $197$~ms shown in Fig. \ref{fig:Rb78resonance} (b). The excitation energy of the isomer when using $^{78n}$Rb$^{8+}$ as the reference is $108.7(6.4)$~keV. When a joint fit is performed on the data, taking into account the correlations between measurements, the energy of the isomer relative to the ground state is $108.4(7.6)$~keV. When time correlations are neglected, the excitation energy of the isomer is $108.7(7.8)$~keV. The results are summarized in Table \ref{tab:energylevel}. Both are in agreement with the more precise value of $111.19(22)$~keV obtained from $\gamma$-ray spectroscopy \cite{NNDC}. It is unsurprising that the excitation energy derived from using $^{78n}$Rb$^{8+}$ as the reference is more precise since fewer systematic effects enter into the calculation. It is interesting to note that the uncertainty on the energy difference, before systematic effects are included, for both the case with time correlations and with $^{78n}$Rb$^{8+}$ as the reference are identical.

\section{Conclusion}

The ground and isomeric state in $^{78}$Rb have been fully resolved in TITAN's precision mass measurement Penning trap through the use of highly charged ions. This is the first time a full separation of the resonance shapes, for this isotope and isomer, could be achieved in a Penning trap. The mass excess values extracted for both states agree with the AME2003 values, as does the measured excitation energy of the isomer. The uncertainty on the weighted ground state mass excess has been reduced by a factor of 2 compared to AME03, resulting in a value of $-66935.3(3.6)$~keV. This is a proof-of-principle that highly charged ions are a powerful tool for increasing the precision and resolution of Penning trap mass spectrometers for on-line spectroscopy and searches for isomeric states.

\begin{acknowledgments} This work has been supported by the Natural Sciences and Engineering Research Council of Canada and the National Research Council of Canada. The authors would like to thank P.M. Walker for his discussions and the TRIUMF technical staff, especially M. Good. T.B. acknowledges support from the Evangelisches Studienwerk e.V. Villigst, S.E. from the Vanier CGS, and V.V.S. from the Deutsche Studienstiftung.
\end{acknowledgments}

%

\end{document}